\documentclass[11pt]{article}
\usepackage{moriond,epsfig}

\def\Journal#1#2#3#4{{#1} {\bf #2}, #3 (#4)}

\def\PRB{{\em Phys. Rev.} B}

\def\be{\begin{equation}}
\def\ee{\end{equation}}
\def\bea{\begin{eqnarray}}
\def\eea{\end{eqnarray}}

\begin{document}
\title{TRANSPORT OF INTERACTING ELECTRONS THROUGH A PAIR OF COUPLED METALLIC 
QUANTUM DOTS}
\author{\underline{ANDREI D. ZAIKIN}$^{1,3}$,
DMITRI S. GOLUBEV$^{2,3}$}

\address{$^1$ Forschungszentrum Karlsruhe, Institut f\"ur Nanotechnologie,
76021, Karlsruhe, Germany}
\address{$^2$Institut f\"ur Theoretische Festk\"orperphysik,
Universit\"at Karlsruhe, 76128 Karlsruhe, Germany}
\address{$^3$I.E. Tamm Department of Theoretical Physics, 
P.N. Lebedev Physics Institute, 119991 Moscow, Russia }

\maketitle\abstracts{We derive a complete expression for the interaction
  correction to the $I-V$ curve of two connected in series metallic quantum
dots. For strongly asymmetric dots in a wide range of parameters this 
interaction correction depends logarithmically on voltage and temperature. 
}

\section{Introduction}

Recently it was demonstrated \cite{GZ04} that important information about
the effect of electron-electron interactions on quantum transport in disordered
conductors can be obtained within a transparent theoretical framework 
of quasiclassical Langevin equations generalized to situations in which
relaxation of the electron distribution function occurs at much
longer time scales as compared to the electron dwell time between
two neighboring scatterers. The main goal of this paper is to employ
the approach \cite{GZ04} in order to study interaction effects in low
temperature electron transport through the system of three quantum scatterers
or, equivalently, two coupled quantum dots. This structure can serve as a
model for a number of transport experiments. In addition to that, the
system of three scatterers appears to be the simplest one in which 
the interaction correction to the $I-V$ curve is described by two 
different contributions, one of them being absent in a yet simpler system 
of two quantum scatterers \cite{GZ03}. 

\section{General Results}

Let us briefly recollect our results \cite{GZ04} 
for the electron-electron interaction correction to the current 
in a chain of $N$ arbitrary coherent scatterers (with resistances $R_n$)
or $N-1$ quantum dots.
Assuming that the dimensionless conductances of the scatterers 
$g_n=2\pi/e^2R_n=2\sum_kT_k^{(n)}$ are large (here and below $T_k^{(n)}$
denotes the transmission of the $k$-th conducting mode in the 
$n$-th scatterer), $g_n \gg 1$, one can effectively expand the general
expression for the current in the inverse conductance and find
\begin{equation}
I=\frac{V}{R_\Sigma}+\delta I_1+\delta I_2, \label{If}
\end{equation}
where \cite{GZ04} 
\begin{eqnarray}
\delta I_1&=&\frac{1}{4eR_\Sigma}\sum_{n,m=1}^N g_m \int
dtdt'\frac{T^2\sin eVt}{\sinh^2\pi Tt}
K_{mn}(t-t') \big[\delta_{m}(a_{m-1}-a_{n-1})D_{n-1,m}(t')-\delta_{m-1}
\nonumber\\ &&
\times(a_{m}-a_{n-1})D_{n-1,m-1}(t')
+\delta_{m}(a_{m-1}-a_{n})D_{n,m}(t')
-\delta_{m-1}(a_{m}-a_{n})D_{n,m-1}(t')\big]
\label{dI2}
\end{eqnarray}
\begin{eqnarray}
\delta I_2&=&-\frac{\pi}{eR_\Sigma^2}\sum_{n,m=1}^N \beta_nR_n\int
dtdt'\frac{T^2\sin eVt}{\sinh^2\pi Tt}
K_{mn}(t-t')\bigg[\delta_{nm}\delta(t')
-\frac{g_m}{4\pi}\big(\delta_{m-1}D_{n-1,m-1}(t')
\nonumber\\ &&
-\,\delta_{m}D_{n-1,m}(t')\big)-\frac{g_m}{4\pi}\big(\delta_{m}D_{nm}(t')
-\delta_{m-1}D_{n,m-1}(t')\big) \bigg]. 
\label{dI1}
\end{eqnarray}
Here $V$ is the voltage bias applied to the chain of quantum dots, 
$R_\Sigma=\sum_{n=1}^N R_n$ is the total chain resistance, $\beta_n=
\sum_kT_k^{(n)}(1-T_k^{(n)})/\sum_kT_k^{(n)}$ is the Fano factor of the
$n$-th scatterer, $\delta_n$ stands for the mean level spacing
in the $n$-th dot $\delta_n$ and $a_n=\sum_{j=1}^n R_j/R_\Sigma$.  

The diffuson $D_{n,m}(t)$ satisfies the equation
\begin{eqnarray}
\frac{\partial D_{n,m}}{\partial
t}&=&\frac{\delta_n}{4\pi}\big(g_nD_{n-1,m}+g_{n+1}D_{n+1,m}
-(g_n+g_{n+1})D_{n,m}\big)+\delta_{nm}\delta(t)
\label{Deq}
\end{eqnarray}
with the boundary conditions
\begin{equation}
D_{0,m}=D_{N,m}=D_{n,0}=D_{n,N}=0. \label{Dbound}
\end{equation}
The function $K_{nm}(t)$ is defined by the formula
\begin{equation}
K_{nm}(t)=e^2\int \frac{d\omega}{2\pi}\,{\rm e}^{-i\omega t}\,\frac{Z_{nm}(\omega)}{-i\omega +0},
\end{equation}
where the effective impedance $Z_{nm}(\omega)$ is determined by the following
set of equations
\begin{eqnarray}
\delta q_n&=&C_{n+1}Z_{n+1,m}-C_nZ_{nm}-C_{gn}\sum_{j=1}^n Z_{jm}
\nonumber\\
-i\omega \delta q_n&=&\frac{Z_{nm}}{R_n}-\frac{Z_{n+1,m}}{R_{n+1}}+
\frac{g_n\delta_{n-1}}{4\pi}\delta
q_{n-1}+\frac{g_{n+1}\delta_{n+1}}{4\pi}\delta q_{n+1}
\nonumber\\
&-&\frac{(g_n+g_{n+1})\delta_{n}}{4\pi}\delta q_n -\delta_{nm}+\delta_{n+1,m},
\nonumber\\
\sum_{n=1}^N Z_{nm}&=&0.
\label{Zeq}
\end{eqnarray}
Here we introduced the capacitances of the $n$-th scatterer and the $n$-th
dot, respectively $C_n$ and $C_{gn}$, as well as the Kronecker 
symbols $\delta_{nm}$.

\section{Two quantum dots}

In Ref. 1 we have explored the above general results for a chain of
$N-1$ identical quantum dots connected by $N$ identical scatterers. In that
case Eqs. (\ref{Deq}), (\ref{Zeq}) can be resolved for arbitrary number of
scatterers $N$. Here we restrict the number of scatterers to $N=3$,
i.e. consider transport of interacting electrons through a pair of coupled
quantum dots. As we have already pointed out, this system appears to be the
simplest one in which both interaction corrections $\delta I_1$ and $\delta
I_2$ differ from zero. Below we will establish explicit analytical expressions 
for both these corrections and analyze them in several important limiting
cases.

In the case of a chain with only three scatterers $D_{nm}(t)$ is described by 
a $2\times 2$ matrix.  Defining the inverse electron dwell times for both
quantum dots, $1/\tau_1=(g_1+g_2)\delta_1/4\pi$ and
$1/\tau_2=(g_2+g_3)\delta_2/4\pi$, and performing the Fourier transformation
one arrives at the equation
\begin{equation}
\left(
\begin{array}{cc}
-i\omega + 1/\tau_1 & -\delta_1 g_2/4\pi \\
-\delta_2 g_2/4\pi & -i\omega+1/\tau_2
\end{array}
\right)
\left(
\begin{array}{cc}
D_{11} & D_{12} \\
D_{21} & D_{22}
\end{array}
\right)=\left(
\begin{array}{cc}
1 & 0 \\
0 & 1
\end{array}
\right),
\end{equation}
which can be easily resolved. Analogously, the impedance matrix $Z_{nm}$ has a
$3\times 3$ structure and it can be directly evaluated as well. Substituting 
the
corresponding results into Eqs. (\ref{dI1}), (\ref{dI2}) we obtain 
\begin{eqnarray}
\delta I_1&=&-2eT\frac{R_1R_2R_3}{R_\Sigma^3}
\frac{W\left(\frac{\gamma_2+ieV}{2\pi T}\right)-W\left(\frac{\gamma_1+ieV}{2\pi T}\right)}
{\sqrt{1-\frac{4\delta_1\delta_2R_1R_2R_3R_\Sigma}
{[\delta_1R_2R_3+\delta_2R_1R_2+(\delta_1+\delta_2)R_1R_3]^2}}},
\label{di2}\\
\delta I_2&=&-\frac{2eT}{R^2_\Sigma(C_{\Sigma 1}C_{\Sigma 2}-C_2^2)}\bigg\{
{\cal B}_1
\frac{W\left(\frac{\kappa_2+ieV}{2\pi T}\right)-W\left(\frac{\kappa_1+ieV}{2\pi T}\right)}{\kappa_2-\kappa_1}
\nonumber\\ &&
+{\cal B}_2
\bigg[\frac{\kappa_2W\left(\frac{\kappa_1+ieV}{2\pi T}\right)-\kappa_1W\left(\frac{\kappa_2+ieV}{2\pi T}\right)}
{\kappa_1\kappa_2(\kappa_2-\kappa_1)}-\frac{\gamma_2W\left(\frac{\gamma_2+ieV}{2\pi T}\right)
-\gamma_1W\left(\frac{\gamma_1+ieV}{2\pi T}\right)}{\kappa_1\kappa_2(\gamma_2-\gamma_1)}\bigg]
\nonumber\\ &&
+{\cal B}_3
\frac{W\left(\frac{\gamma_2+ieV}{2\pi T}\right)-W\left(\frac{\gamma_1+ieV}{2\pi T}\right)}{\kappa_1\kappa_2(\gamma_2-\gamma_1)}
\bigg\},
\label{di1}
\end{eqnarray}
Here we defined $W(x)={\rm Im}[x\Psi(1+x)]$, where $\Psi(x)$ is the digamma
function. We also defined 
$C_{\Sigma 1}=C_1+C_2+C_{g1}$, $C_{\Sigma 2}=C_2+C_3+C_{g2}$ as well as 
\begin{eqnarray}
{\cal B}_1&=&C_{\Sigma 1}\beta_1R_1+C_{\Sigma 2}\beta_3R_3+(C_{\Sigma 1}+C_{\Sigma 2}-2C_2)\beta_2R_2,
\nonumber\\
{\cal B}_2&=&\beta_1\frac{R_1(R_2+R_3)}{R_2R_3}+\beta_2\frac{R_2(R_1+R_3)}{R_1R_3}+\beta_3\frac{R_3(R_1+R_2)}{R_1R_2},
\nonumber\\
{\cal B}_3&=&\left(\left(\frac{1}{R_2}+\frac{1}{R_3}\right)\frac{1}{\tau_2}+\frac{\delta_1g_2}{4\pi}\frac{1}{R_2}\right)\beta_1R_1
\nonumber\\ &&
+\,\left(\left(\frac{1}{R_1}+\frac{1}{R_2}\right)\frac{1}{\tau_1}+\frac{\delta_2g_2}{4\pi}\frac{1}{R_2}\right)\beta_3R_3
+\left(\frac{1}{R_1}\frac{\delta_1g_1}{4\pi}+\frac{1}{R_3}\frac{\delta_2g_3}{4\pi}\right)\beta_2R_2
\nonumber
\end{eqnarray}
and
\begin{eqnarray}
\kappa_{1,2}&=&\frac{\nu
\mp\sqrt{\nu^2
-4(C_{\Sigma 1}C_{\Sigma 2}-C_2^2)
\left(\frac{1}{R_1R_2}+\frac{1}{R_1R_3}+\frac{1}{R_2R_3}\right)
}}{2(C_{\Sigma 1}C_{\Sigma 2}-C_2^2)},
\nonumber\\
\nu&=&\frac{C_{\Sigma 1}}{R_3}+\frac{C_{\Sigma 2}}{R_1}+\frac{C_{\Sigma 1}+C_{\Sigma 2}-2C_2}{R_2},
\nonumber\\
\gamma_{1,2}&=&\frac{1}{2}\left(\frac{1}{\tau_1}+\frac{1}{\tau_2}\right)
\mp \frac{1}{2}\sqrt{\left(\frac{1}{\tau_1}-\frac{1}{\tau_2}\right)^2+\frac{\delta_1\delta_2g_2^2}{4\pi^2}}.
\nonumber
\end{eqnarray}

The above equations fully determine the leading interaction correction to the
current in a system of two coupled quantum dots. These equations can now be
analyzed in various physical limits. For the case of identical scatterers and
dots our results match with those derived in Ref. 1, while in the limit
$R_2=0$ the problem is reduced to that of a single quantum dot \cite{GZ03}. 

Let us first restrict our attention to the case of fully 
open quantum dots $\beta_{1,2,3}=0$. In this case the term (\ref{di1}) 
vanishes identically, $\delta I_2\equiv 0$, and the effect of
electron-electron interactions on the $I-V$ curve is described only by the
$\beta$-independent term $\delta I_1$ (\ref{di2}). At $T\to 0$ for the
differential conductance we obtain 
\begin{eqnarray}
\frac{dI}{dV}=\frac{1}{R_\Sigma}-\frac{e^2}{2\pi}
\frac{R_1R_2R_3}
{R_\Sigma^3\sqrt{1-\frac{4\delta_1\delta_2R_1R_2R_3R_\Sigma}
{[\delta_1R_2R_3+\delta_2R_1R_2+(\delta_1+\delta_2)R_1R_3]^2}}}
\ln\frac{\gamma_2^2+e^2V^2}{\gamma_1^2+e^2V^2}.
\label{intcorr}
\end{eqnarray}
This equation demonstrates that at large voltages $eV \gg \gamma_2$ the 
interaction correction remains small while for $eV < \gamma_1$ it saturates
to a finite voltage-independent value, i.e. at such voltages the $I-V$ curve
of our system is Ohmic. In the limit $\gamma_1 \ll \gamma_2$ the interaction
correction depends logarithmically on voltage provided the condition   
 $\gamma_1 \ll eV \ll \gamma_2$ is fulfilled. 

Assume now that one of the two
dots, e.g. the second one, is very large. In this case both $\delta_2$ and
$1/\tau_2$ remain small. Setting $\delta_2 \to 0$, for $eV\tau_1 \ll 1$
from (\ref{intcorr}) one finds
\begin{eqnarray}
\frac{dI}{dV}=\frac{1}{R_\Sigma}\left(1+\frac{2}{g_\Sigma}
\frac{R_1R_2R_3}{R_\Sigma^3}\ln (eV\tau_1)\right),
\label{intcorr2}
\end{eqnarray}
where $g_\Sigma =2\pi/e^2R_\Sigma$. For $\delta_2 = 0$ this result remains
valid down to exponentially low voltages.

\begin{figure}
\begin{center}
\includegraphics[width=11cm]{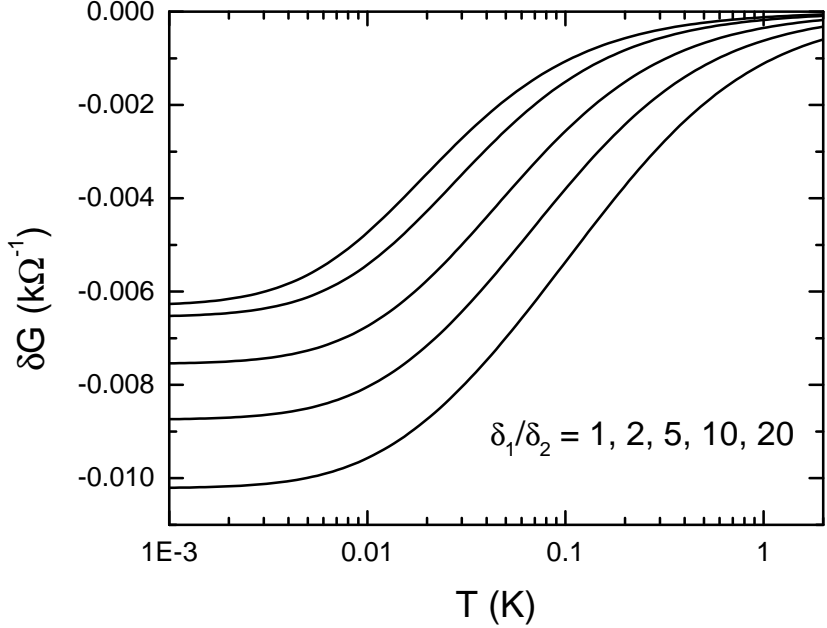}
\end{center}
\caption{The temperature dependent interaction correction $\delta G (T)$ to
  the linear conductance for a pair of open quantum dots ($\beta_{1,2,3}=0$)
  with $R_{1,2,3}= 1$ $\Omega$ and $\delta_2=1.5$ mK. The ratio $\delta_1/\delta_2$ increases from top to
  bottom.}
\end{figure}

Now let us briefly consider a general case of partially transparent 
quantum dots $\beta_{1,2,3} \neq 0$. In this case the term $\delta I_2$
(\ref{di1}) differs from zero and should be added to the correction $\delta
I_1$ studied above. Provided the asymmetry between two dots is not large, 
at low enough voltages $eV < \gamma_1$ the term $\delta_1$ saturates to a
finite value similarly to $\delta I_1$. For $\delta_2 \to 0$ and 
$eV\tau_1 \ll 1$ the correction $\delta_1$ depends logarithmically on voltage
(cf. Eq. (\ref{intcorr2})).
Adding both $\delta I_1$ and $\delta I_2$ together and assuming 
$R_3 \ll R_1,R_2$, for $eV\tau_1 \ll 1$ and $T \to 0$ we obtain
\begin{eqnarray}
\frac{dI}{dV}=\frac{1}{R_1+R_2}\left(1+\frac{2\tilde \beta_{12}}{g_3}\ln
(eV\tau_1)\right)+\delta G,
\label{intcorr3}
\end{eqnarray}
where $\delta G$ is the negative voltage-independent contribution 
(not presented here) originating from energies above $1/\tau_1$ and 
$$
\tilde\beta_{12}=
\frac{R_1^2(R_2+R_1\beta_1)+R_2^2(R_1+R_2\beta_2)}{(R_1+R_2)^3}
$$ 
is the total Fano factor \cite{GZ04,GZ03} for
the first quantum dot. The result (\ref{intcorr3}) matches with one obtained in Ref. 2 for
a single quantum dot shunted by an Ohmic resistor. In our case the role of
such a shunt is played by a large quantum dot with $\delta_2 \to 0$.

\section*{Acknowledgments}

This work is part of the
Kompetenznetz ``Funktionelle Nanostructuren'' supported by the Landestiftung
Baden-W\"urttemberg gGmbH and of the STReP project ``Ultra-1D'' supported by
the EU.

\end{document}